\documentclass[12pt]{article}

\usepackage{epsfig}

\def\square{\kern1pt\vbox{\hrule height 1.2pt
\hbox{\vrule width 1.2pt\hskip 3pt
\vbox{\vskip 6pt}\hskip 3pt\vrule width 0.6pt}
\hrule height 0.6pt}\kern1pt}
\def\ltwid{\mathrel{\raise.3ex\hbox{$<$\kern-.75em\lower1ex\hbox{$\sim$}}}}

\begin{document}

\begin{titlepage}
\begin{flushright}
CRETE-09-11 \\ UFIFT-QG-09-02
\end{flushright}

\vspace{0.5cm}

\begin{center}
\bf{A Phenomenological Model for the Early Universe}
\end{center}

\vspace{0.3cm}

\begin{center}
N. C. Tsamis$^{\dagger}$
\end{center}
\begin{center}
\it{Department of Physics, University of Crete \\
GR-710 03 Heraklion, HELLAS.}
\end{center}

\vspace{0.2cm}

\begin{center}
R. P. Woodard$^{\ast}$
\end{center}
\begin{center}
\it{Department of Physics, University of Florida \\
Gainesville, FL 32611, UNITED STATES.}
\end{center}

\vspace{0.3cm}

\begin{center}
ABSTRACT
\end{center}
\hspace{0.3cm} We consider the description of cosmological
dynamics from the onset of inflation by a perfect fluid 
whose parameters must be consistent with the strength of 
the enhanced quantum loop effects that can arise during 
inflation. The source of these effects must be non-local 
and a simple incarnation of it is studied both analytically 
and numerically. The resulting evolution stops inflation 
in a calculable amount of time and leads to an oscillatory 
universe with a vanishing mean value for the curvature 
scalar and an oscillation frequency which we compute. 

\vspace{0.3cm}

\begin{flushleft}
PACS numbers: 04.60.-m, 04.62.+v, 98.80.Cq
\end{flushleft}

\vspace{0.1cm}

\begin{flushleft}
$^{\dagger}$ e-mail: tsamis@physics.uoc.gr \\
$^{\ast}$ e-mail: woodard@phys.ufl.edu
\end{flushleft}

\end{titlepage}

\section{Introduction}

Although it is not yet known how to account for the
late time acceleration of the universe \cite{riess,wang}, 
it is by now quite clear that an adequate period of 
approximately exponential expansion -- inflation 
\cite{weinberg} -- provides a simple and natural 
explanation for the homogeneity and isotropy of the 
large-scale observable universe \cite{bernardis}, 
and also is in satisfactory agreement with the 
primordial density perturbations spectrum \cite{dunkley}.
This inflationary phase is usually realized by a scalar
field but this is not a necessity.

Consider the pure gravitational equations of motion:
\footnote{Hellenic indices take on spacetime values while 
Latin indices take on space values. The Hubble constant 
is $3H^2_0 \equiv \Lambda$. Our metric tensor has
spacelike signature and our curvature tensor equals:
$R^{\alpha}_{~\beta\mu\nu} \equiv 
\Gamma^{\alpha}_{~\nu\beta, \mu} +
\Gamma^{\alpha}_{~\mu\rho} \;
\Gamma^{\rho}_{~\nu\beta} -
(\mu \leftrightarrow \nu)$.} 
\begin{equation}
G_{\mu\nu} \; \equiv \;
R_{\mu\nu} \, - \, \frac12 g_{\mu\nu} \, R \; = \;
- \Lambda \, g_{\mu\nu}
\;\; . \label{eom1}
\end{equation}
If $\Lambda$ is assumed to be positive, the ``no-hair''
theorems imply that classically the local geometry 
approaches the maximally symmetric solution at late 
times \cite{AbbDes}; this solution is de Sitter 
spacetime and, thus, $\Lambda$-driven inflation is 
intrinsic to (\ref{eom1}). 

As  (\ref{eom1}) shows, classical gravitation without 
matter is a theory which only ``knows'' about the 
cosmological constant $\Lambda \,$; Newton's constant $G$ 
sets the strength of quantum effects. The corresponding
mass scales are the Planck mass $M_{\rm Pl}$ -- associated 
with $G$ -- and the mass $M$ -- associated with $\Lambda$:
\begin{equation}
M^2_{\rm Pl} \; \equiv \; \frac{1}{G} 
\qquad , \qquad
M \; \equiv \; \Big( \frac{\Lambda}{8 \pi G} \Big)^\frac14
\;\; . \label{scales} 
\end{equation}
We restrict ourselves to scales below the Planck mass
so that the dimensionless coupling constant $\epsilon 
\equiv G\Lambda$ of the theory is small:
\footnote{This is a very mild restriction on the range 
of scales; for instance, if $M \sim 10^{16} \, GeV$ we get 
that $\epsilon \equiv G\Lambda \sim \frac{M^4}{M^4_{\rm Pl}}
\sim 10^{-12}$.}
\begin{equation}
M \; < \; M_{\rm Pl}
\qquad \Longleftrightarrow \qquad
\epsilon \, \equiv \, G\Lambda \; < \; 1
\;\; . \label{epsilon} 
\end{equation}

The quantum behaviour of gravity in de Sitter ($dS$)
spacetime has been studied in perturbation theory and in
the infrared \cite{NctRpw1a, NctRpw1b}. In particular, 
the expansion rate $H(t)$ decreases by an amount which 
becomes non-perturbatively large at late times:
\footnote{The Hubble parameter $H(t)$ shall be defined
in Section 2.}
\begin{equation}
de \; Sitter
\quad \Longrightarrow \quad
H(t) \; \simeq \;
H_0 \left\{ 1 \, - \, 
\epsilon^2 \Big[ \, \# (H_0 \, t) + O(1) \, \Big] \, + \,
O(\epsilon^3) \right\}
\;\; , \label{HdS}
\end{equation}
where $\#$ is a positive pure number of $O(1)$.
The underlying physical mechanism is the production of
infrared quanta out of the vacuum due to the accelerated
expansion of spacetime. Such a production can only occur
for particles that are light compared to the Hubble scale 
without classical conformal invariance; gravitons and 
massless minimally coupled scalars are unique in that 
respect.

The factor of $H_0 \, t$ which appears in expression (4) 
is known as an {\it infrared logarithm} because it derives 
from infrared virtual particles and because $H_0 \, t$ is 
the logarithm of the de Sitter scale factor. Any quantum 
field theory which involves undifferentiated gravitons or 
massless minimally coupled scalars will show infrared 
logarithms in some correlators at some order in the loop 
expansion. If the interaction contains $N$ undifferentiated 
gravitons or massless minimally coupled scalars, along with 
any number of other fields, then each new factor of the 
coupling constant squared can produce at most $N$ additional 
infrared logarithms. For example, the fundamental interaction 
of quantum gravity in de Sitter background takes the generic 
form \cite{NctRpw2}:
\begin{equation}
\sqrt{G} \; h \, \partial h \, \partial h
\;\; , \label{dSint}
\end{equation}
where $h_{\mu\nu}$ is the fluctuating graviton field. Thus, 
one can get at most one extra infrared logarithm for each 
additional power of $G$ \cite{NctRpw1b}. 

The operator under study also has an effect. For example, 
because there are two derivatives in the invariant measure 
of acceleration \cite{NctRpw3} whose expectation value gave 
expression (4), the general form of such corrections is:
\begin{equation}
H(t) \; = \;
H_0 \, \Biggl\{
1 \, - \, \sum_{\ell=2}^{\infty} \epsilon^{\ell} \,
\sum_{k=0}^{\ell-1} c_{\ell k} \,
(H_0 \, t)^k \Biggr\} 
\;\; , \label{Hgen}
\end{equation}
where $\ell$ stands for the loop order and where the 
constants $c_{\ell k}$ are pure numbers of $O(1)$.

Because $\epsilon$ is constant, whereas $H_0 \, t$ grows 
without bound, infrared logarithms eventually lead to 
a breakdown of perturbation theory. In quantum gravity 
on de Sitter background this occurs after about 
$\epsilon^{-1}$ e-foldings. To evolve further requires 
a non-perturbative technique such as summing the series 
of leading infrared logarithms:
\begin{equation}
H(t) \Big\vert_{\rm leading \; log} \; = \; 
H_0 \, \Biggl\{
1 \, - \, 
\epsilon \sum_{\ell=2}^{\infty}
c_{\ell, \, \ell-1} \; (\epsilon H_0 \, t)^{\ell-1} 
\Biggr\} 
\;\; . \label{Hgenll}
\end{equation}
Starobinski\u{\i} \cite{AAS} has developed a stochastic 
technique which exactly reproduces the leading infrared 
logarithms of scalar potential models \cite{NctRpw1b}, 
and which can be used to sum them whenever the scalar 
potential is bounded below \cite{SY}. Starobinski\u{\i}'s 
technique has recently been extended to include models 
in which the scalar interacts with other fields such as 
a Yukawa fermion \cite{MW1} or electromagentism \cite{PTW}. 
The late time limits of the vacuum energies of these scalar 
models exhibit a broad range of possibilities for what the 
quantum gravitational sum might give:
\begin{itemize}
\item{Scalar potential models which are bounded below 
show a small, constant increase of the vacuum energy 
\cite{SY,SR};}
\item{Scalar quantum electrodynamics experiences a small, 
constant decrease of the vacuum energy \cite{PTW}; and}
\item{Yukawa theory engenders a decrease of the vacuum 
energy which grows without bound \cite{MW1}.}
\end{itemize}

We would ultimately like to {\it compute} how quantum 
gravity affects late time cosmology by employing 
Starobinski\u{\i}'s technique to sum the series of 
leading infrared logarithms. There has been some progress 
in this area \cite{MW2} but the full solution is not yet 
in sight. A more modest approach is to anticipate the 
solution by attempting to guess the most cosmologically 
significant part of the effective field equations guided 
by our understanding of the perturbative regime at leading 
logarithm order. The resulting model could be regarded as 
a worthy object of study in its own right, just as one views 
the many classes of scalar-driven inflation models, without 
feeling any need to derive them from fundamental theory. 
It may even be that our study will uncover some general 
feature of any successful model which can, it turn, guide 
the fundamental derivation.

It should be noted that doubts have been raised about the 
possibility of {\it any} infrared contribution to the 
quantum gravitational vacuum energy \cite{GT,NctRpw4}. However, 
these doublts are difficult to reconcile with the fact that 
scalar models certainly show infrared corrections to the 
vacuum energy \cite{SY,SR,PTW,MW1}, and with Weinberg's 
observation that infrared logarithms contaminate the power 
spectrum of scalar-driven inflation \cite{SW}. General 
theoretical arguments have also been advanced to show that 
de Sitter must be unstable in quantum gravity \cite{AMP}.

One feature which complicates evaluation of these arguments 
is the intractibility of quantum gravity at any order, and 
the fact that the onset of this particular effect occurs 
at two loops. The latter fact must be so because screening 
represents the gravitational attraction between virtual 
infrared gravitons which have been ripped from the vacuum. 
The production process is a one-loop effect, so the 
gravitational response to it cannot occur until the next 
loop order. Three separate computations of the graviton 
1-point function have confirmed that there is no one-loop 
effect \cite{Ford,FMVV,NctRpw5}, and the same conclusion can 
be reached from taking the de Sitter limit of scalar-driven 
inflation \cite{JM,JS}.
\footnote{Note that the slow roll suppression one finds 
for corrections to the background in this limit merely 
means that some of the fields must be differentiated, 
as in the $\sqrt{G} \; h \, \partial h \, \partial h$ 
vertex. Hence, one can only get a single extra infrared 
logarithm for each extra $G$, as opposed to the three 
infrared logarithms that would be possible if the vertex 
had been $\sqrt{G} \; h^3$.} This difficulty of performing 
explicit computations is one more reason why it might 
be desirable to study quantum gravitational screening 
from the perspective of the effective field equations.

In the present paper we shall use the physical principles 
responsible for the non-trivial quantum gravitational 
back-reaction on inflation to construct a phenomenological 
model which we can then directly evolve. Therefore, we wish 
to construct an appropriate {\it effective} conserved 
stress-energy tensor $T_{\mu\nu}[g]$ which will modify 
the gravitational equations of motion (\ref{eom1}) in 
the usual way:
\begin{equation}
G_{\mu\nu} \; = \;
- \Lambda \, g_{\mu\nu} \, + \,
8\pi G \, T_{\mu\nu}[g]
\;\; . \label{eom2}
\end{equation}
Our stress-energy tensor must be a {\it non-local} functional 
of the metric as dictated by the nature of the effect we 
wish to describe. It can be conveniently parametrized as 
a ``perfect fluid'': 
\begin{equation}
T_{\mu\nu}[g] \; = \;
(\rho + p) \, u_{\mu} \, u_{\nu} \, + \,
p \, g_{\mu\nu}
\;\; , \label{Tmn}
\end{equation}
so that to completely determine it we need the following
three ingredients: \\
{\it (i)} the energy density $\rho$ as a functional
of the metric tensor $\rho[g](x)$, \\
{\it(ii)} the pressure $p$ as a functional of the
metric tensor $p[g](x)$, \\
{\it (iii)} the 4-velocity field $u_{\mu}$ as a
functional of the metric tensor $u_{\mu}[g](x)$,
chosen to be timelike and normalized: 
\begin{equation}
g^{\mu\nu} \, u_{\mu} u_{\nu} = -1
\qquad \Longrightarrow \qquad
u^{\mu} \, u_{\mu ; \nu} = 0
\;\; . \label{u}
\end{equation}

Section 2 describes a simple {\it ansatz} for the effective
stress-energy tensor $T_{\mu\nu}[g]$ and the set of equations 
it leads to for homogeneous and isotropic spacetimes. Section 
3 presents numerical results obtained by discretizing the 
relevant evolution equation. To the degree that is possible, 
the dynamics of homogeneous and isotropic evolution in the 
presence of $T_{\mu\nu}[g]$ is studied analytically in Section 
4 and -- wherever a comparison can be made -- the excellent 
agreement with our numerical study is noticed. Section 5 
discusses late time evolution and the possible modifications 
it implies on our {\it ansatz}. Our conclusions comprise 
Section 6.

\section{A Physical Ansatz}

It is clearly impossible to uniquely fix the functional 
form of the effective stress-energy tensor solely from 
the physical requirements and correspondence limits we 
have at our disposal; only quantum field theory could, 
in principle, provide such an answer. However, we can 
try to obtain a {\it simple} ansatz and then analyze 
its implications. \\ 

{$\bullet \;\; $} {\bf Why a Perfect Fluid?} \\
One might think that a good way of discussing quantum 
corrections to the field equations would be in terms
of an ansatz for the effective action. However, the
``in-out'' effective action -- being non-local -- does 
not give causal effective field equations we can use 
to study cosmological evolution. The ``in-in'' effective
action of the Schwinger-Keldysh formalism does produce 
causal effective field equations, but this results from 
subtle cancellations between different off-shell fields, 
making it difficult to identify promising candidates 
for the off-shell effective action. For a class of simple 
non-local effective actions, it is possible -- using a
partial integration ``trick'' -- to obtain causal 
effective field equations \cite{deser} but it is 
impossible to restrict the non-local effects to the 
past. The chain of action and reaction that is part 
of any single-field variational formalism implies that 
screening the cosmological constant by an inverse 
differential operator acting on some curvature scalar 
will inevitably lead to a variation of that curvature 
scalar appearing {\it at the current time} in the 
effective field equations. Hence one always gets a 
renormalization of the effective Newton constant. 
We wish to avoid this, so we specify the effective 
field equations directly and insist that the non-local 
screening of the cosmological constant both remains 
in the distant past and also does not change the 
Einstein tensor so that Newton's constant is not
renormalized. The problem with this procedure is 
that we must enforce conservation. We selected the 
perfect fluid form for the effective stress tensor 
both because enforcing conservation is straightforward 
and because our perturbative studies of quantum 
gravitational screening indicate that this form 
suffices to capture the leading infrared logarithms
we seek to reproduce \cite{NctRpw6}. \\

{$\bullet \;\; $} {\bf Implications of Conservation} \\
The ``perfect fluid'' parametrization (\ref{Tmn}) of 
$T_{\mu\nu}[g]$ allows us to completely determine it
from the three quantities it contains: the scalars 
$\rho$, $p$ and the 4-vector  $u_{\mu}$. Because of 
the normalization (\ref{u}), only three of the 
components of $u_{\mu}$ are algebraically independent.
Thus, $T_{\mu\nu}[g]$ contains five independent 
quantities in total. Conservation provides four 
equations and allows us to determine any four of 
these quantities in terms of any one. It turns out 
to be more convenient to specify the induced pressure 
functional $p[g]$ and then use conservation to obtain 
the form of the induced energy density $\rho[g]$ and 
4-velocity $u_{\mu}[g]$ up to their initial value data.

The fundamental equation:
\begin{equation}
D^{\mu} \, T_{\mu\nu} \; = \; 0
\;\; , \label{cons1}
\end{equation}
implies:
\begin{equation}
\partial_{\nu} \, p 
\, + \,
u_{\nu} \, ( \, u \cdot D + D \cdot u \, ) (\rho + p)
\, + \,
(\rho + p) ( \, u \cdot Du_{\nu} \, ) 
\; = \; 0
\;\; . \label{cons2} 
\end{equation}
By contracting $u^{\nu}$ into (\ref{cons2}) and 
using (\ref{u}), we get: 
\begin{eqnarray}
u \cdot \partial \, p \!& = &\!
D_{\mu} \Big[ \, (\rho + p) u^{\mu} \, \Big]
\;\; , \label{cons3a} \\
u \cdot \partial \, \rho \!& = &\! 
- (D \cdot u) (\rho + p)
\;\; . \label{cons3b}
\end{eqnarray}
Then, by substituting (\ref{cons3b}) into the 
conservation equation (\ref{cons2}), the following 
equation emerges:  
\begin{equation}
(\rho + p) \; u \cdot D u_{\nu} \; = \;
- ( \, \partial_{\nu} \, + \, 
u_{\nu} \, u \cdot \partial \, ) \, p
\;\; . \label{cons4}
\end{equation}
Equations (\ref{cons3a}-\ref{cons4}) can be used
to accomplish our goal. \\

{$\bullet \;\; $} {\bf Requirements on the Pressure} \\

\vspace{-0.4cm}

{\it (i) The initial value requirement} \\
Our gravitationally induced source should not disturb the 
basic nature of the pure gravitational equations (\ref{eom1}). 
The latter can be evolved from the initial spacelike surface 
knowing only the metric and its first time derivative.
This property of gravity must be retained in the presence 
of the source and constrains both the local and non-local
parts of its functional form \cite{NctRpw7}; for instance,
any local parts in $T_{\mu\nu}[g]$ can contain at most 
second time derivatives of the metric. \\
 
{\it (ii) The non-locality requirement} \\
We argued that the physical effect responsible for 
gravitationally inducing $T_{\mu\nu}[g]$ is inherently
non-local and, therefore, our source must be non-local.
It is important to mention that this conclusion can 
also be reached by noting that no local modification
of pure gravity can prevent de Sitter spacetime from 
being a solution of the field equations eternally 
\cite{NctRpw7}. Any local modification simply changes 
the initial Hubble constant $H_0$ and can be absorbed 
by the cosmological constant counterterm $\delta \Lambda$
to leave no change and de Sitter spacetime as a solution
for all time. Thus, the important part of the induced 
stress-energy tensor {\it must} be non-local. \\
 
{\it (iii) The simplicity requirement} \\
A simple non-local operator at our disposal is 
the inverse of the scalar d'Alembertian:
\footnote{Our scalar d'Alembertian is defined with 
retarded boundary conditions.}
\begin{equation}
\square \, \equiv \;
\frac{1}{\sqrt{-g}} \;
\partial_{\mu} \Big( \,
g^{\mu\nu} \sqrt{-g} \; \partial_{\nu} \, \Big)
\;\; , \label{box}
\end{equation}
and a simple scalar it can act on is the curvature 
scalar $R$. Hence, we shall explore {\it ansatze} 
in which the pressure is a function of the quantity
$X[g]$:
\begin{equation}
X \; \equiv \; 
\frac{1}{\square} \, R
\;\; . \label{X}
\end{equation}  

{\it (iv) The correspondence requirement} \\
Our gravitationally induced source should reproduce 
the perturbative results obtained in de Sitter
spacetime. \\

{$\bullet \;\; $} {\bf Cosmological Spacetimes} \\
The large-scale homogeneity and isotropy of the universe
selects Friedman-Robertson-Walker ($FRW$) spacetimes as 
those of primary cosmological interest; their line element
for zero spatial curvature equals in co-moving coordinates:
\begin{equation}
ds^2 \; = \; 
g_{\mu\nu}(t) \; dx^{\mu} dx^{\nu} \; = \; 
- dt^2 \, + \, a^2(t) \; d{\vec x} \cdot d{\vec x} 
\;\; . \label{ds^2frw}
\end{equation}   
Derivatives of the scale factor $a(t)$ give the Hubble 
parameter $H(t)$ -- a measure of the cosmic expansion rate 
-- and the deceleration parameter $q(t)$ -- a measure of 
the cosmic acceleration:
\begin{eqnarray}
H(t) & \equiv &
\frac{{\dot a}(t)}{a(t)} \, = \,
\frac{d}{dt} \ln a(t)
\;\; , \label{H} \\
q(t) & \equiv &
- \frac{a(t) \, {\ddot a}(t)}{{\dot a}^2(t)} \, = \,
-1 \, - \, \frac{\dot H(t)}{H^2(t)}
\;\; . \label{q}
\end{eqnarray} 

For these spacetimes the stress-energy tensor (\ref{Tmn}) 
takes the form: 
\begin{eqnarray}
T_{00} & = & 
u_0 \, u_0 \, ( \rho + p ) \, - \, p 
\; = \;
\rho
\;\; , \label{T00frw} \\
T_{0i} & = & 0
\;\; , \label{T0ifrw} \\
T_{ij} & = & 
u_i \, u_j \, ( \rho + p ) \, + \, 
g_{ij} \, p 
\; = \;
g_{ij} \, p
\;\; . \label{Tijfrw}
\end{eqnarray} 
An immediate consequence of isotropy and the normalization
condition (\ref{u}) is: 
\begin{equation}
u_{\mu} \; = \;
- \, \delta_{\mu}^{~0} 
\qquad \Longleftrightarrow \qquad
u^{\mu} \; = \;
\delta^{\mu}_{~0} 
\;\; . \label{umfrw}
\end{equation}
The Ricci tensor and Ricci scalar become, respectively:
\begin{eqnarray}
R_{00} & = & 
- \left[ \, \frac{3{\ddot a}}{a} \, \right] 
\; = \; 
- \left( \, 3H^2 \, + \, 3 {\dot H} \, \right)
\;\; , \label{R00frw} \\
R_{0i} & = & 
0
\;\; , \label{R0ifrw} \\
R_{ij} & = & 
\left[ \, \frac{{\ddot a}}{a} \, + \,
\frac{2{\dot a}^2}{a^2} \, \right] g_{ij} 
\; = \; 
\left( \, 3H^2 \, + \, {\dot H} \, \right) g_{ij}
\;\; , \label{Rijfrw} 
\end{eqnarray}
and:
\begin{equation}
R \; = \;
\left[ \, \frac{6{\ddot a}}{a} \, + \,
\frac{6{\dot a}^2}{a^2} \, \right] 
\; = \; 
\left( \, 12H^2 \, + \, 6 {\dot H} \, \right)
\;\; . \label{Rfrw}
\end{equation}

In view of (\ref{T00frw}-\ref{Tijfrw},
\ref{R00frw}-\ref{Rfrw}), the non-trivial gravitational 
equations of motion (\ref{eom2}) take the form:
\begin{eqnarray}
3H^2 & = &
\Lambda \, + \, 8 \pi G \, \rho
\;\; , \label{g00frw} \\
-2 {\dot H} - 3 H^2 & = & \!
- \Lambda \, + \, 8 \pi G \, p
\;\; ,\label{gijfrw}
\end{eqnarray}
while the conservation equation (\ref{cons1}) becomes:
\footnote{The $FRW$ conservation equation (\ref{consfrw})
can also be derived directly from the equations of motion 
(\ref{g00frw}, \ref{gijfrw}).}
\begin{equation}
\dot{\rho} \; = \;
- 3H \, ( \rho + p ) 
\;\; . \label{consfrw}
\end{equation}
The latter implies that:
\begin{equation}
\rho (t) \; = \; 
- p(t) \, + \,
\frac{1}{a^3(t)} 
\int_0^t dt' \; a^3(t') \; \dot{p} (t') 
\;\; . \label{rhofrw-pfrw} \\
\end{equation}

When acting on functions which only depend on co-moving
time, the scalar d'Alembertian (\ref{box}) for $FRW$ 
geometries equals:
\begin{equation}
\square \; = \;
- \left( \, \partial_t^2 \, + \, 3H \partial_t \, \right)
\;\; , \label{boxfrw}
\end{equation}
so that its inverse is:
\begin{equation}
\frac{1}{\square} \; = \;
- \int_0^t dt' \; \frac{1}{a^3(t')} \, 
\int_0^{t'} dt'' \; a^3(t'')
\;\; . \label{invboxfrw} 
\end{equation}
Consequently, the source $X$ can be written as follows:
\begin{equation}
X \; = \;
\frac{1}{\square} \, R \; = \;
- \int_0^t dt' \; \frac{1}{a^3(t')} \,
\int_0^{t'} dt'' \; a^3(t'') \,
\left[ \, 12 H^2(t'') \, + \, 6 {\dot H}^2(t'') \, \right]
\;\; . \label{Xfrw} 
\end{equation}
Note that we have taken the initial time to be at 
$t=0$. \\ 

{$\bullet \;\; $} {\bf The de Sitter Correspondence Limit} \\
If we define inflation as positive expansion, i.e. 
$H(t) > 0$, with negative decelaration, i.e. $q(t) < 0$,
a locally de Sitter geometry provides the simplest
paradigm. It is characterized by constant Hubble and 
deceleration parameters, and a scale factor of a simple 
exponential form:
\begin{equation}
H_{dS}(t) \; = \; H_0
\qquad , \qquad
q_{dS}(t) \; = \; -1
\qquad , \qquad
a_{dS}(t) \; = \; e^{H_0 t}
\;\; . \label{dS1}
\end{equation}
The source $X_{dS}$ can be computed by first obtaining
the inverse d'Alembertian and curvature scalar from
(\ref{invboxfrw}) and (\ref{Rfrw}) respectively:
\begin{eqnarray}
\frac{1}{\square} \, \Big\vert_{dS} & = &
- \int_0^t dt' \; e^{-3 H_0 t'} \, 
\int_0^{t'} dt'' \; e^{3 H_0 t''}
\;\; , \label{invboxdS} \\
R_{dS} & = & 12H^2
\;\; , \label{RdS}
\end{eqnarray}
and then acting the former on the latter:
\begin{eqnarray}
X_{dS} \; = \;
\Big( \, \frac{1}{\square} \, R \, \Big)_{dS} \!\!& = &\!\!
- \int_0^t dt' \; e^{-3 H_0 t'} \, 
\int_0^{t'} dt'' \; e^{3 H_0 t''} \;
12H^2
\nonumber \\
\!\!& = &\!\!
-4 H_0 \int_0^t dt' \; \left[ \, 1 - e^{-3 H_0 t'} \, \right]
\nonumber \\
\!\!& = &\!\!
-4 H_0 \, t \; + \; 
\frac43 \left[ \, 1 - e^{-3 H_0 t} \, \right]
\;\; . \label{XdS}
\end{eqnarray}
For large observation times we get:
\begin{equation}
X_{dS} \; \simeq \;
-4 \ln [ \, a_{dS}(t) \, ] \, + \, O(1)
\qquad , \qquad
\ln [ \, a_{dS}(t) \, ] = H_0 t
\;\; . \label{XdS2}
\end{equation}
>From (\ref{XdS2}) we already see that our source $X$ 
can give the infrared logarithm in our perturbative 
result (\ref{HdS}). We should mention at this stage
that explicit perturbative computations for a massless
minimally coupled self-interacting scalar field give
the following leading logarithm result \cite{OW}:
\begin{equation}
leading \; log 
\quad \Longrightarrow \quad
H^2(t) \, = \,
\frac{\Lambda}{3} \, \Bigg\{
1 \, - \, G \Lambda \, 
\sum_{\ell = 2}^{+\infty} \, h_{\ell} \,
\Big[ \, G \Lambda \, \ln [ \, a_{dS}(t) \, ] 
\, \Big]^{\ell - 1} \Bigg\}
\label{HdS2}
\end{equation}
where $\ell$ is the loop order and $h_{\ell}$ are
pure numbers. Now consider the equation of motion 
(\ref{gijfrw}) for de Sitter spacetime:
\footnote{According to our perturbative result 
(\ref{HdS}), ${\dot H}_{dS}(t)$ is subdominant since
its time derivative must eliminate one infrared 
logarithm $\ln [a_{dS}(t)]$ without affecting the
corresponding factor of $\epsilon \equiv G \Lambda$.} 
\begin{equation}
-3H^2 \; \simeq \; -3H_0^2 \, + \,
8 \pi G \, p[g_{dS}]
\;\; . \label{gijdS}
\end{equation}
The leading logarithm form of the induced pressure
immediately follows:
\begin{equation}
leading \; log 
\quad \Longrightarrow \quad
p(t) \, = \,
\Lambda^2 \, 
\sum_{\ell = 2}^{+\infty} \, h_{\ell} \,
\Big[ \, G \Lambda \, \ln [ \, a_{dS}(t) \, ] 
\, \Big]^{\ell - 1} 
\;\; . \label{pdS2}
\end{equation}

In view of the above, the following {\it ansatz} for 
the gravitationally induced pressure $p[g_{dS}](t)$ 
reproduces (\ref{HdS}) up to a numerical coefficient:
\begin{equation}
p[g_{dS}](t) \; = \;
\Lambda^2 \, f[- \epsilon \, X_{dS}](t) \; = \;
\Lambda^2 \, f[- (G\Lambda) (\square^{-1} R)_{dS}](t)
\;\; , \label{pdS}
\end{equation}
where $f$ is some monotonically unbounded function.
The reason for requiring $f$ to be unbounded is 
dictated by the similar behaviour seen in our lowest 
order perturbative result (\ref{HdS}). Moreover, 
since the source $X_{dS}$ is increasingly negative 
definite as can be seen from (\ref{XdS}), any function
$f$ satisfying: 
\begin{equation}
f[- \epsilon \, X_{dS}] \; = \;
- \epsilon \, X_{dS} \, + \,
O(\epsilon^2)
\;\; , \label{fincrdS}
\end{equation}
will be increasingly positive definite. 

Finally, successively apply equations (\ref{pdS}),
(\ref{fincrdS}) and (\ref{XdS2}) to the equation of
motion (\ref{gijdS}):
\begin{eqnarray}
H^2 \!\!\!& \simeq &\!\!\! 
H_0^2 \, \Big\{ \, 1 \, - \, 8 \pi \epsilon \, 
f[- \epsilon \, X_{dS}] \, \Big\}
\nonumber \\
\!\!\!& \simeq &\!\!\!
H_0^2 \, \Big\{ \, 1 \, - \, 8 \pi \epsilon \, \left[ \, 
- \epsilon \, X_{dS} + O(\epsilon^2) \, \right] \Big\}
\nonumber \\
\!\!\!& \simeq &\!\!\!
H_0^2 \, \Big\{ \, 1 \, - \, 32 \pi \epsilon^2 \, 
\ln (a_{dS}) \, + \, O(\epsilon^3) \, \Big\}
\;\; . \label{gijdS2}
\end{eqnarray}
Up to a positive numerical coefficient, 
\footnote{In \cite{NctRpw6}, the pure number $h_2$ was
calculated and was found to equal $\frac{1}{12 \pi^3}$.
Since this was done within the context of a simplified 
quantum gravitational theory, we shall instead assume 
it equals one and, therefore, it does not alter the
coefficient of the $- \epsilon X_{dS}$ term in 
(\ref{fincrdS}).}
the requisite agreement is achieved. \\

{$\bullet \;\; $} {\bf The General Ansatz} \\
Our physical requirements and correspondence limits have
led us to the following {\it ansatz} for the gravitationally
induced pressure $p[g](x)$ in a general geometry:
\begin{equation}
p[g](x) \; = \;
\Lambda^2 \, f[- \epsilon \, X](x) 
\qquad , \qquad
X \, \equiv \, \frac{1}{\square} \, R
\;\; , \label{p}
\end{equation}
where the function $f$ satisfies:
\begin{equation}
f[- \epsilon \, X] \; = \;
- \epsilon \, X \, + \, O(\epsilon^2)
\;\; . \label{fincr}
\end{equation}
To completely determine the induced stress-energy tensor 
$T_{\mu\nu}[g](x)$ we need the energy density $\rho[g](x)$ 
and the 4-velocity $u_{\mu}[g](x)$. Given the pressure 
$p[g](x)$, we can obtain the other two quantities via 
stress-energy conservation up to their initial value data.

An explicit cosmological model needs an explicit 
function $f$. Out of the plethora of functions satisfying
(\ref{fincr}) we shall select {\it simple} ones and 
define:
\begin{eqnarray}
& (i) &
{\rm the} \,\, linear \,\, {\rm model}: \qquad
f[- \epsilon \, X] \; \equiv \;
- \epsilon \, X 
\;\; , \label{linear} \\
& (ii) & 
{\rm the} \,\, exponential \,\, {\rm model}: \qquad
f[- \epsilon \, X] \; \equiv \;
e^{- \epsilon \, X} \, - \, 1
\;\; . \qquad \label{exp}
\end{eqnarray} 
It is the dynamical evolution of the explicit cosmological 
models that will determine whether, as we expect, they will 
naturally stop inflation. Afterwards, the induced source
should ``turn-off'' and the universe should enter a 
radiation dominated epoch.

\section{Numerical Results}

Because the non-local, non-linear equations we have 
proposed are too complicated to solve exactly, we 
shall evolve them numerically. For that purpose, it 
is preferable to use the $(ij)$ equation of motion 
due to its linearity in the highest derivative:
\begin{equation}
2 {\dot H} + 3 H^2 \; = \;
3 H_0^2 \, \Big\{ \, 1 \, - \,
8 \pi \epsilon \, f[ -\epsilon \, X ] 
\, \Big\}
\qquad , \qquad
X \; \equiv \; 
\frac{1}{\square} \, R
\;\; , \label{evoeqn1}
\end{equation}
where, as we have already mentioned, $H_0$ is the 
Hubble parameter at the onset of inflation and 
$\epsilon \equiv G \Lambda = 3 G H_0^2$ is the 
dimensionless coupling constant of the theory.  

The discretization of (\ref{evoeqn1}) involves: \\
{\it (i) Constants,}
\begin{eqnarray}
{\rm step \; size \; in \; Hubble \; units}
& \Longrightarrow &\quad
\delta \; \equiv \;
H_0 \, \Delta t
\;\; , \label{delta} \\
{\rm coupling \; constant}
&\Longrightarrow &\quad
\epsilon \; \equiv \;
G \Lambda 
\;\; , \label{couplconst}
\end{eqnarray}
{\it (ii) The basic variables and initial value data,}
\footnote{It is of course equivalent to 
consider $(\rho + p)$ instead of $p$ because
of stress-energy conservation; see equation 
(\ref{consfrw}).}
\begin{eqnarray}
a(t)
& \longrightarrow &\quad
a(i \, \Delta t) \; = \; 
e^{b_i}
\qquad , \qquad
b_0 = 0
\;\; , \label{bi} \\
\Big( \rho + p \Big)(t)
& \longrightarrow &\quad
[\rho + p](i \, \Delta t) 
\qquad , \qquad 
[\rho + p \Big]_0 = 0
\;\; , \nonumber \\
& \mbox{} &
[\rho + p \Big]_{i+1} \; = \;
e^{-3 b_i} \Big\{ \,
[\rho + p]_i - \epsilon \, \Delta X_i \,
f'[ - \epsilon \, X_i ] \, \Big\} 
\; . \qquad \label{rho+pi}
\end{eqnarray}
All quantities of interest as well as their initial 
values can be determined from (\ref{bi}-\ref{rho+pi}): \\
{\it (i) Dynamical quantities,}
\begin{eqnarray}
H(t)
& \longrightarrow &\quad
H(i \, \Delta t) \; = \; 
\frac{\Delta b_i}{\Delta t}
\qquad , \qquad 
\Delta b_i \; \equiv \;
b_{i+1} \, - \, b_i
\;\; , \label{Hi} \\
{\dot H}(t)
& \longrightarrow &\quad
{\dot H}(i \, \Delta t) \; = \; 
\frac{\Delta^2 b_i}{\Delta t^{\, 2}}
\qquad , \qquad
\Delta^2 b_i \; \equiv \;
\Delta b_{i+1} \, - \, \Delta b_i
\;\; , \qquad \label{dotHi} \\
q(t)
& \longrightarrow &\quad
q(i \, \Delta t) \; = \; 
- \, 1 - \frac{\Delta^2 b_i}{\Delta b_i^{\, 2}}
\;\; , \qquad \label{qi} \\
R(t) 
& \longrightarrow &\quad
R (i \; \Delta t) \; = \;
\frac{1}{\Delta t^2} \,
\Big[ \, 6 \Delta^2 b_i + 12 (\Delta b_i)^2
\, \Big] 
\;\; , \label{Ri} \\
X(t)
& \longrightarrow &\quad
X(i \, \Delta t)
\quad , \quad
X_{i+1} \; = \;
X_i + \Delta X_i
\;\; , \nonumber \\
& \mbox{} &
\hspace{0.5cm}
\Delta X_{i+1} \; = \;
e^{-3 b_i} \Big[ \,
\Delta X_i - 12 \, (\Delta b_i)^2 - 
6 \, \Delta^2 b_i \, \Big] 
\;\; , \label{Xi}
\end{eqnarray}
{\it (ii) Initial value data,}
\begin{eqnarray}
& \mbox{} &
\Delta b_0 = \delta
\qquad , \qquad 
\Delta^2 b_0 = 0 
\;\; , \label{ivdbi} \\
& \mbox{} &
X_0 = \Delta X_0 = 0
\;\; . \label{ivdXi}
\end{eqnarray}

The discretized evolution equation (\ref{evoeqn1})
reads:
\begin{equation}
\Delta^2 b_i \; = \; 
\frac32 \Big[ \,  \delta^2 - (\Delta b_i)^2 \, \Big]
\, - \,
12 \pi \delta^2 \epsilon \, f[ -\epsilon \, X_i ] 
\;\; . \label{evoeqni} 
\end{equation}
It was numerically integrated using {\it Mathematica}
for the following choice of the input parameters and
of the step range:
\begin{equation}
\delta \; = \; \frac{1}{1000}
\quad , \quad
\epsilon \; = \;
\frac{1}{200}
\qquad ; \qquad
i \in [0, 350000]
\;\; . \label{input1}
\end{equation}
Moreover, for the function $f$ we chose the one
corresponding to the exponential model (\ref{exp}):
\begin{equation}
f(x) \; = \; e^{x} - 1
\quad \Longrightarrow \qquad
f^{-1}(x) \; = \; \ln (1+x)
\quad , \quad
f'(x) \; = \; e^{x}
\;\; . \label{input2}
\end{equation}
The associated critical point $x_{cr} = 
- \epsilon X_{cr}$ -- defined in (\ref{Xcr}) --
and frequency $\omega$ -- defined in (\ref{omega}) 
-- are:
\begin{eqnarray}
X_{cr} \!\!& = &\!\!
- \, \frac{1}{\epsilon} \,
f^{-1} \Big( \frac{1}{8 \pi \epsilon} \Big) 
\; = \; 
- \, \frac{1}{\epsilon} \,
\ln \Big( 1 + \frac{1}{8 \pi \epsilon} \Big)
\; \sim \; - \, 438.50
\;\; , \label{Xcr2} \\
\omega \!\!& = &\!\!
\frac{\epsilon \, \delta}{\Delta t} \, 
\sqrt{72 \pi \, f'_{cr}}
\; = \;
\frac{\epsilon \, \delta}{\Delta t}  \, 
\Bigg[ \, 72 \pi \, 
\Big( 1 + \frac{1}{8 \pi \epsilon} \Big)
\, \Bigg]^{\frac12}
\; \sim \;
\frac{2.25 \times 10^{-4}}{\Delta t}
\;\; . \qquad \label{omega2}
\end{eqnarray}

Our results are presented in the set of graphs
that can be found in the very end. Some comments 
are in order: \\
{$\bullet \; $} After the onset of and during the
era of inflation, the source $X(t)$ grows while
the curvature scalar $R(t)$ and Hubble parameter
$H(t)$ decrease. \\
{$\bullet \; $} Inflation ends and the time when 
this occurs is the time when the deceleration 
parameter $q(t)$ goes from negative to positive 
values. \\
{$\bullet \; $} During the era of oscillations: \\
{\it (i)} The oscillations of $R(t)$ are centered
around $R = 0$, have an envelope behaving like
$t^{-1}$ and a frequency $\omega$ in agreement 
with (\ref{omega2}) \\
{\it (ii)} Although there is net expansion, the 
oscillations of $H(t)$ take it to small negative 
values for small time intervals. The presence of 
these short deflation periods is a novel feature 
of the model and may have consequences on the 
primordial perturbation spectrum. \\   
{\it (iii)} The oscillations of $\dot{H}(t)$ show
that there is almost no difference between $R(t)$
and $6 \dot{H}$ and, therefore, the term in $R(t)$
proportional to $H^2(t)$ is insignificant during
this era. \\
{\it (iv)} The oscillations of the scale factor
$a(t)$ are centered around a linear increase with
time. 

It is important to note at this stage the
excellent agreement of the analytical results
derived in Section 4 with their numerical
equivalents. The two basic parameters to
concentrate are, \\
{$\bullet \; $} {\it Criticality:} It occurs 
at step $i = 160942$, as the detailed data of 
Figure 12 indicates, and at that point:
\begin{equation}
X[160942] \; = \; -438.50
\qquad , \qquad
q[160942] \; = \; 0.50
\;\; , \label{numcrit}
\end{equation}  
in complete agreement with the analytical 
predictions (\ref{Xcr2}) and (\ref{qcr}) 
respectively. \\
{$\bullet $} {\it Oscillation frequency:} 
>From the detailed data of Figure 4 we conclude 
that six oscillations have occured between steps 
$i = 174291$ and $i = 342478$. Hence, we have:
\begin{equation}
T \; = \; 
\frac{342478 - 174291}{6} \, \Delta t
\; = \; 
\frac{2 \pi}{\omega}
\qquad \Longrightarrow \qquad
\omega \; = \; 
\frac{2.24 \times 10^{-4}}{\Delta t}
\;\; , \label{numomega}
\end{equation}
which compares very well with the analytical 
prediction (\ref{omega2}).

\section{Analytical Results}

With the evolution equation (\ref{evoeqn1}) as a
starting point, we can analytically derive some
results for the physical system under study. Our 
{\it ansatz} restricts the function $f$ to be
monotonically unbounded. Therefore, there exists 
a critical point $X_{cr}$ such that:
\begin{equation}
1 \, - \, 8 \pi \epsilon \, f[ -\epsilon \, X_{cr} ]
\; = \; 0
\qquad \Longrightarrow \qquad
X_{cr} \; = \;
- \, \frac{1}{\epsilon} \;
f^{-1} \Big( \frac{1}{8 \pi \epsilon} \Big)
\;\; . \label{Xcr}
\end{equation}
Inflationary evolution dominates roughly until we  
reach the critical point. Close to the critical
point the induced pressure $p$ is small and, thus,
it makes sense to expand $f$ around its critical 
point and use the resulting perturbation theory
for the subsequent evolution:
\begin{eqnarray} 
2 {\dot H} + 3 H^2 \!\!& = &\!\!
3 H_0^2 \, \Big\{ \, 1 \, - \,
8 \pi \epsilon \, 
f[ -\epsilon \, X_{cr} - 
\epsilon (X - X_{cr}) ] \, \Big\}
\nonumber \\
& = &\!\!
3 H_0^2 \, \Bigg\{ \, 1 \, - \,
8 \pi \epsilon \, \Big( \, 
f[ -\epsilon \, X_{cr} ] \, - \,
\epsilon (X - X_{cr}) \, 
f'[ -\epsilon \, X_{cr} ] \, \Big)
\qquad \nonumber \\ 
& \mbox{} &
\hspace{3cm}
+ \, O \Big( \epsilon^2 (X - X_{cr})^2 \Big)
\; \Bigg\}
\nonumber \\
& \simeq &\!\!
24 \pi \epsilon^2 \, H_0^2 \,
(X - X_{cr}) \,
f'[ -\epsilon \, X_{cr} ]
\;\; . \label{evoeqn2}
\end{eqnarray}
Neglecting all higher order terms is a superb 
approximation given the very small realistic
values of $\epsilon$.

Moreover, using (\ref{Rfrw}) we rewrite the co-moving
time derivative of the Hubble parameter as:
\begin{equation}
{\dot H} \; = \;
\frac16 R \, - \, 2H^2
\;\; . \label{Hdot}
\end{equation}
Consequently, the evolution equation (\ref{evoeqn2})
becomes:
\begin{equation}
-R \, + \, 3 H^2 \; \simeq \;
-72 \pi \, ( \epsilon H_0 )^2 \,
(X - X_{cr}) \, f'_{cr} 
\;\; , \label{evoeqn3}
\end{equation}
where we have defined:
\begin{equation}
f'_{cr} \; \equiv \;
f'[ -\epsilon \, X_{cr} ] \; \equiv \;
- \frac{1}{\epsilon} \,
\frac{d}{dX} f[ -\epsilon \, X ] 
\Big\vert_{X = X_{cr}}
\;\; . \label{f'cr}
\end{equation}
Action of the d'Alembertian operator (\ref{boxfrw})
on (\ref{evoeqn3}) gives:
\begin{equation}
\ddot{R} \, + \, 2H \, \dot{R} \, + \,
( \omega^2 - {\dot H} )\, R \, + \,
\Big[ \, 3H^2 R - 36H^4 \, \Big] 
\; \simeq \; 0
\;\; , \label{evoeqn4}
\end{equation}
with the understanding that:
\begin{equation}
\omega \; \equiv \; 
\epsilon H_0 \,
\sqrt{72 \pi \, f'_{cr}}
\;\; . \label{omega}
\end{equation}
It will turn out that the term in brackets is 
subdominant and we can focus our attention to 
the differential equation:
\begin{equation}
\ddot{R} \, + \, 2H \, \dot{R} \, + \,
\Big( \omega^2 - {\dot H} \Big) \, R
\; \simeq \; 0
\;\; , \label{evoeqn5}
\end{equation}
which describes a damped oscillator. To solve 
the above equation we first scale out the Hubble 
friction term by defining:
\begin{equation}
R \; \equiv \;
\frac{1}{a} \, S
\;\; , \label{S}
\end{equation}
so that (\ref{evoeqn5}) becomes:
\begin{equation}
\ddot{S} \, + \, 
\Big( \omega^2 - 2{\dot H} - H^2 \Big) \, S
\; \simeq \; 0
\;\; , \label{evoeqn6}
\end{equation}
and, in the large time limit, is solved by:
\begin{equation}
S(t) \; \simeq \;
K_1 \, \sin( \omega t + \varphi ) 
\qquad , \qquad
\omega^2 \; \gg \;
\Big\vert - 2{\dot H} - H^2 \Big\vert
\;\; . \label{evoeqnS}
\end{equation}
Hence, by using (\ref{S}) and (\ref{Rfrw}), 
we have:
\begin{equation}
K_1 \, \sin( \omega t + \varphi ) \; \simeq \;
a \, R \; = \;
6 \, \frac{d}{dt} ( H a ) \, + \,
6 H^2 a \; \simeq \;
6 \, \frac{d}{dt} ( H a )
\;\; . \label{evoeqnR}
\end{equation}
>From (\ref{evoeqnR}) we immediately conclude:
\begin{eqnarray}
{\dot a}(t) \!\!& \simeq &\!\!
K_2 \, - \, \frac{K_1}{6 \, \omega} \,
\cos( \omega t + \varphi )
\;\; , \label{evoeqndota} \\
a(t) \!\!& \simeq &\!\!
K_3 \, + \, K_2 \, t \, - \,
\frac{K_1}{6 \, \omega^2} \,
\sin( \omega t + \varphi )
\;\; . \label{evoeqna}
\end{eqnarray}
and our large time results are:
\footnote{These results justify our ignoring the
bracketed term in equation (\ref{evoeqn4}) and the 
$H^2 a $ term in equation (\ref{evoeqnR}).}
\begin{eqnarray}
\lim_{t \gg 1} \, H(t) \!\!& \simeq &\!
\frac{1}{t} \, - \,
\frac{K_1}{6 K_2 \, \omega} \, 
\frac{\cos( \omega t + \varphi )}{t} \, + \,
O \Big( \frac{1}{t^2} \Big)
\;\; , \label{Hlarget} \\
\lim_{t \gg 1} \, {\dot H}(t) \!\!& \simeq &\!
\frac{K_1}{6 K_2} \, 
\frac{\sin( \omega t + \varphi )}{t} \, + \,
O \Big( \frac{1}{t^2} \Big)
\;\; , \label{Hdotlarget} \\
\lim_{t \gg 1} \, R(t) \!\!& \simeq &\!
\frac{K_1}{K_2} \, 
\frac{\sin( \omega t + \varphi )}{t} \, + \,
O \Big( \frac{1}{t^2} \Big)
\;\; . \label{Rlarget}
\end{eqnarray}
>From the asymptotic solution that we just obtained,
the physical picture that emerges so far is that 
of a universe in which, {\it as the exit from the 
inflationary era approaches, oscillations in $R$ 
become significant}; their frequency $\omega$ is 
given by (\ref{omega}) and, according to 
(\ref{Rlarget}), their envelope is linearly 
falling with time.

We can get further insight by computing the
deceleration parameter $q$ at criticality.
This is most easily done by considering the
equations of motion (\ref{g00frw}-\ref{gijfrw})
from which we deduce that:
\begin{equation}
-2 {\dot H} \; = \;
8 \pi G \, ( \rho + p )
\; = \;
\Lambda \, 8 \pi \epsilon \, 
\frac{\rho + p}{\Lambda^2}
\;\; . \label{Hdotfrw1} 
\end{equation}
Then, we rewrite (\ref{g00frw}) in a form convenient
for our purpose:
\begin{eqnarray}
3H^2 \!\!& = &\!\!
\Lambda \, + \, 8 \pi G \, \rho
\nonumber \\
& = &\!\!
\Lambda \, \Big[ \, 1 \, + \, 
8 \pi \epsilon \, \frac{\rho}{\Lambda^2} \, \Big]
\; = \; 
\Lambda \, \Big[ \, 1 \, - \, 
8 \pi \epsilon \, \frac{p}{\Lambda^2} \, + \,
8 \pi \epsilon \, \frac{\rho + p}{\Lambda^2} 
\, \Big]
\;\; , \label{g00frw2} 
\end{eqnarray}
and use (\ref{Hdotfrw1}-\ref{g00frw2}) to express 
the decelaration parameter (\ref{q}) in the following
way:
\begin{equation}
q \; = \;
-1 \, - \, \frac{\dot H}{H^2}
\; = \;
-1 \; + \; \frac32 \times
\frac
{ \Lambda \, 8 \pi \epsilon \, 
\frac{\rho + p}{\Lambda^2} }
{ \Lambda \, \Big[ \, 1 \, - \, 
8 \pi \epsilon \, \frac{p}{\Lambda^2} \, + \,
8 \pi \epsilon \, \frac{\rho + p}{\Lambda^2} 
\, \Big] }
\;\; . \label{q2}
\end{equation}
The definitions (\ref{Xcr}) of criticality and 
(\ref{p}) of pressure imply:
\begin{equation}
1 \, - \, 8 \pi \epsilon \, f[ -\epsilon \, X_{cr} ]
\; = \; 
1 \, - \, 8 \pi \epsilon \, \frac{p_{cr}}{\Lambda^2}
\; = \; 0
\;\; , \label{pcr}
\end{equation}
and allow us to conclude that:
\begin{equation}
q_{cr} \; = \; -1 + \frac32 \; = \; + \, \frac12
\;\; \label{qcr}
\end{equation}
At the onset of inflation $q_0 = -1$. Since by the 
time the universe arrived at the critical point the 
decelaration parameter had already reached positive 
values --  $q_{cr} = + \frac12$ -- {\it the epoch 
of inflation ended before the universe evolved to
the critical time}.

To investigate whether the dynamical system is
underdamped at the critical point, we isolate all 
terms of the full equation (\ref{evoeqn4}) that
affect the frequency:
\begin{equation}
frequency \;\; terms
\qquad \Longrightarrow \qquad
\Big( \, \omega^2 - {\dot H} + 3H^2 \, \Big) \, R
\;\; . \label{oscfreq}
\end{equation}
We have already seen that:
\begin{equation}
q_{cr} \; = \;
-1 \, - \, \frac{\dot H_{cr}}{H_{cr}^2} \; = \;
+ \, \frac12
\qquad \Longrightarrow \qquad
- {\dot H}_{cr} \; = \;
\frac32 \, H_{cr}^2
\;\; , \label{Hcr}
\end{equation}
leading to a positive frequency determining 
coefficient:
\begin{equation}
\Big( \, \omega^2 - {\dot H} + 3H^2 \, \Big) \,
\Big\vert_{cr} \; = \;
\omega^2 \, + \, \frac92 \, H_{cr}^2 
\, > \, 0
\;\; . \label{freqcr}
\end{equation}
Hence, at criticality the system is underdamped 
implying again that {\it oscillations start 
around the end of inflation}.

It is also interesting to work out an approximate
but direct relation between the frequency $\omega$ 
and the Hubble parameter $H$ at the critical time.
The starting point is equation (\ref{Hdotfrw1})
and the approximation consists of evaluating its
right hand side in de Sitter spacetime:
\begin{equation}
-2 {\dot H} \; \sim \;
\frac{8 \pi \epsilon}{\Lambda} \, ( \rho + p ) 
\Big\vert_{dS} \; = \;
32 \pi \, (\epsilon H_0)^2 \, f'_{dS}
\;\; , \label{HdotfrwdS1}
\end{equation}
and afterwards at criticality:
\begin{equation}
{\dot H}_{cr} \; \sim \;
-16 \pi \, (\epsilon H_0)^2 \, f'_{dS}
\;\; . \label{HdotfrwdS2}
\end{equation}
Comparison of (\ref{HdotfrwdS2}) with (\ref{omega})
gives us the desired approximate expressions:
\begin{equation}
\frac{{\dot H}_{cr}}{\omega^2} \; \sim \;
- \, \frac{6}{27}
\qquad , \qquad
\frac{H^2_{cr}}{\omega^2} \; \sim \;
+ \, \frac{4}{27}
\;\; , \label{Hcr-omega}
\end{equation}
where we have used (\ref{Hcr}) to arrive at the 
second relation. Even with the eternal de Sitter
assumption -- which ignores the dimunition of 
the Hubble parameter $H$ with time -- the frequency
$\omega$ is the larger quantity; in the realistic
case the ratios (\ref{Hcr-omega}) would be much
smaller. What we can conclude is that, in our
evolution, the inequality $H^2 < \vert {\dot H} 
\vert < \omega^2$ is well justified. \\

{$\star \; $} {\it Identities} \\
In addition to the equations presented throughout the 
main text, various of the following expressions have 
been used to obtain the results of this subsection: 
\begin{eqnarray}
\dot{R} \!\!& = &\!\!
\frac{1}{a} \Big[ \, \dot{S} - H S \, \Big]
\qquad , \qquad
\ddot{R} \; = \;
\frac{1}{a} \Big[ \, \ddot{S} - 2 H \dot{S} +
H^2 S - \dot{H} S \, \Big]
\;\; . \qquad \label{IDs1}
\end{eqnarray}

\vspace{-0.6cm}

\begin{eqnarray}
X_{dS} \!\!& \simeq &\!\!
-4 H_0 \, t
\qquad , \qquad
{\dot X}_{dS} \; \simeq \;
-4 H_0 
\;\; , \label{IDdS1} \\
( \rho \, + \, p )_{dS}
\!\!& \simeq &\!\!
\frac{1}{3 H_0} \, \dot{p}_{dS}
\; = \;
\frac{1}{3 H_0} \, \Lambda^2
( -\epsilon \, {\dot X}_{dS} ) \, f'_{dS}
\; \simeq \;
12 \, \epsilon \, H_0^4 \, f'_{dS}
\;\; . \qquad \label{IDdS2}
\end{eqnarray}

\section{After Inflation}

The homogeneous and isotropic evolution described 
in the previous two Sections does not give a completely 
satisfactory end to inflation. The oscillations are 
no problem, but the average expansion $a(t) \sim t$ 
is unacceptably rapid. At that rate there would be 
no reheating and the late time universe would be 
cold and empty. Nonetheless, the same is true for 
scalar-driven inflation if one ignores the possibility 
for energy to flow from the inflaton into ordinary 
matter. We believe that energy will flow from the 
gravitational sector of our model into ordinary 
matter to create a radiation-dominated universe, 
just as it is thought to do for scalar-driven 
inflation. In that case, one should think of the
total stress-energy as consisting of our quantum
gravitational perfect fluid plus the energy density
and pressure of radiation, with the latter described
just as in conventional cosmology.

An amazing possibility arises if this process can 
be shown to occur: {\it our quantum gravitational 
correction cancels the bare cosmological constant 
and then becomes dormant during the epoch of radiation 
domination.} To see this, suppose the deceleration 
parameter has the pure radiation value of $q(t) = +1$ 
for times $t > t_r$.  In that case, the Hubble 
parameter and scale factor are:
\begin{eqnarray}
q \, = \, +1 
\quad \Longrightarrow \qquad 
H(t) \!\!& = &\!\! 
\frac{H_r}{1 + 2 H_r (t - t_r)} 
\;\; , \label{radH} \\
a(t) \!\!& = &\!\!
a_r \Big[ \, 1 +  2 H_0 (t - t_r) \, \Bigr]^{\frac12} 
\;\; , \hspace{3cm} \label{rada}
\end{eqnarray}
where $H_r$ and $a_r$ are their values at $t = t_r$. 
During this phase the Ricci scalar is zero:
\begin{equation}
q \, = \, +1 
\quad \Longrightarrow \qquad 
R \; = \; 6 \dot{H} + 12 H^2 \; = \; 0 
\;\; . \label{radR}
\end{equation}
Our simple source $X(t)$ obeys the differential 
equation $\square X = R$, so for $t > t_r$ it must 
be a linear combination of its two homogeneous 
solutions:
\footnote{The equation $\square X = R$ remains 
true even if the  total stress-energy includes 
radiation and/or matter contributions.}
\begin{eqnarray}
\forall \, t \!\!\!& > &\!\!\! t_r
\quad \Longrightarrow \quad
\square X \, = \, 0
\qquad \Longrightarrow 
\nonumber \\
X(t) \!\!& = &\!\!
X_r \, + \,
\dot{X}_r \int^t \!\! dt'
\left[ \frac{a_r}{a(t')} \right]^3 = \,
X_r \, - \, 
\frac{\dot{X}_r}{H_r} \, 
\frac{1}{\sqrt{1 + 2 H_r (t - t_r)}} 
\;\; . \qquad \label{radX}
\end{eqnarray}
The only solution consistent with $q = +1$ is:
\begin{equation}
X_r \; = \; X_{cr} 
\qquad , \qquad
\dot{X}_r \; = \; 0
\;\; . \label{radX2}
\end{equation} 
The system is predisposed to reach nearly $X_{cr}$ 
in any case but one might doubt that evolution would 
enforce the exact vanishing of the second solution 
implied by $\dot{X}_r = 0$. However, note that 
{\it the vanishing of the $\int^t dt' \, a^{-3}(t')$
homogeneous solution is attained} by the purely 
gravitational evolution of Sections 3 and 4, which 
does not include energy transfer to matter. In that 
case $R \neq 0$, so there is a homogeneous contribution 
as well, but the fact that the oscillations are about 
$X_{cr}$ means that the second homogeneous solution 
is completely absent.

Having $X(t)$ approach $X_{cr}$ within the context 
of a hot, radiation dominated universe would be 
a great success for our model, but the eventual 
transition to matter domination poses problems. 
The onset of matter domination is really a gradual 
process but let us simplify the exposition by 
considering a sudden change from $q = +1$ to 
$q = + \frac12$ at some time $t_m \gg t_r$. During 
this matter dominated epoch the Hubble parameter 
and scale factor are:
\begin{eqnarray}
q \, = \, + \, \frac12 
\quad \Longrightarrow \qquad 
H(t) \!\!& = &\!\!
\frac{H_m}{1 + \frac32 H_m (t - t_m)} 
\;\; , \label{matterH} \\
a(t) \!\!& = &\!\!
a_m \Big[ \, 1 + \frac32 H_m (t - t_m) 
\, \Bigr]^{\frac23} 
\;\; , \hspace{3cm} \label{mattera}
\end{eqnarray}
where $H_m$ and $a_m$ are $H(t_m)$ and $a(t_m)$, 
respectively, computed from the radiation dominated 
geometry (\ref{radH}-\ref{rada}). During matter
domination the Ricci scalar is nonzero:
\begin{equation}
q \, = \, + \, \frac12 
\quad \Longrightarrow \qquad 
R \; = \; 6 \dot{H} + 12 H^2 \; = \;
\frac{3 H_m^2}
{\Big[ \, 1 + \frac32 H_m (t - t_m) \, \Big]^2} 
\;\; . \label{matterR}
\end{equation}
The resulting change in the source $X(t)$ is:
\begin{eqnarray}
q \, = \, + \, \frac12 
\;& \Longrightarrow &\quad 
\nonumber \\
\Delta X(t) \!\!\!\!& \equiv &\!\!\!\!
X(t) - X_c \; = \;
-\frac43 \,
\ln\Big[ \, 1 + \frac32 H_m (t - t_m) \, \Big] 
\, + \, O(1) 
\;\; . \qquad \label{matterX}
\end{eqnarray}

To understand what is wrong with the change 
(\ref{matterX}) caused by matter domination, it 
is useful to recall our {\it ansatz} (\ref{p}) 
for the quantum gravitationally induced pressure:
\begin{equation}
p[g](x) \; = \;
\Lambda^2 \, f[- G \Lambda \, X](x) 
\;\; . \label{restate}
\end{equation}
In the context of this ansatz there are two major 
problems with (\ref{matterX}): \\

{$\bullet \; $} {\it The sign problem.} It derives 
from the function $f(x)$ in (\ref{restate}) being 
monotonically increasing and unbounded. Hence, pushing 
$X(t)$ below $X_{cr} \ll 0$ results in positive total 
pressure, whereas observation implies negative pressure 
during the current epoch \cite{riess, wang}. Note that 
we cannot alter this feature of $f(x)$ without 
sacrificing the very desirable ability of the model 
to cancel an arbitrary bare cosmological constant. \\

{$\bullet \; $} {\it The magnitude problem.} In one
sentence, the magnitude of the total pressure produced
by (\ref{matterX}) is vastly too large. The problem 
arises from the factors of the bare cosmological
constant $\Lambda$ in our ansatz (\ref{restate}). 
The total pressure $p_{\rm tot}$ is the sum of the 
classical contribution and our ansatz (\ref{restate}):
\begin{eqnarray}
p_{\rm tot} \!\!& = &\!\!
- \, \frac{\Lambda}{8\pi G} \,
\Biggl\{ \,
1 \, - \, 8 \pi \, G \Lambda \,
f[ \, - G \Lambda \; 
( X_{cr} + \Delta X ) \, ] \, \Biggr\} 
\label{ptot1} \\
\!\!& \simeq &\!\!
- \, \frac{\Lambda}{G} \times 
( G \Lambda )^2 \; f_{cr}' \; \Delta X 
\;\; . \label{ptot2}
\end{eqnarray}
Note that we need not include in $p_{\rm tot}$ an
additional contribution because non-relativistic 
matter has zero pressure. Comparing with the currently 
observed value $p_{\rm now}$ of the pressure:
\begin{equation}  
p_{\rm now} \; \simeq \;
- \, \frac{3}{8 \pi G} \, H_{\rm now}^2
\;\; , \label{pnow}
\end{equation}
gives:
\begin{equation}
\frac{p_{\rm tot}}{p_{\rm now}} 
\; \simeq \;
\left( \frac{G \Lambda \, H_0}{H_{\rm now}} 
\right)^2 f_{cr}' \; \Delta X 
\; \simeq \;
10^{86} \times f_{cr}' \times \Delta X 
\;\; , \label{ratiop}
\end{equation}
where we have assumed $H_0 \sim 10^{13}~{\rm GeV}$ 
and $H_{\rm now} \sim 10^{-33}~{\rm eV}$. The 
derivative $f_{cr}'$ is unity for the linear model 
(\ref{linear}) and of order $(G \Lambda)^{-1} 
\sim 10^{12}$ for the exponential model (\ref{exp}), 
so we expect $f_{cr}'$ to be at least of order 
one and possibly much greater. \\

There is no way of addressing either problem 
without generalizing our ansatz (\ref{restate}) 
for the pressure. This necessarily takes us away 
from what can be motivated by explicit computation 
during the de Sitter regime. Although these issues
will be analyzed elsewhere \cite{NctRpw8}, we 
shall mention the basic principles: \\
{$\bullet$} The magnitude problem arises because 
the constant $\Lambda$ in (\ref{restate}) is about 
the square of the inflationary Hubble parameter 
rather than its late time descendant that could 
be $55$ orders of magnitude smaller. Solving the 
problem entails replacing one of these factors 
of $\Lambda$ present in (\ref{restate}) by some 
dynamical scalar quantity that changes as time 
evolves in a way that also preserves the original 
relaxation mechanism. \\
{$\bullet$} The sign problem arises because the 
Ricci scalar is positive during both inflation 
and matter domination. Again solving the problem 
involves a dynamical scalar quantity that changes
sign from inflation to matter domination and is
still zero during radiation domination.

\section{Epilogue}

We have presented a simple {\it ansatz} for the most 
cosmologically significant part of the effective field 
equations of quantum gravity with a positive cosmological 
constant. The quantum correction to these equations 
consists of a ``perfect fluid'' stress-energy tensor 
in which the pressure is specified as a non-local 
functional of the metric, and the associated energy 
density and timelike 4-velocity are determined by 
conservation. On the basis of simplicity, as well as
correspondence with perturbative results in de Sitter 
background, we proposed that the pressure takes the 
form:
\begin{equation}
p[g](x) \; = \;
\Lambda^2 \, f[ \, - G \Lambda \, 
{\square}^{-1} R \, ](x)
\;\; , \label{restate2}
\end{equation}
where $\square$ is the scalar d'Alembertian and its 
inverse is defined with retarded boundary conditions.

We studied homogeneous and isotropic evolution 
in this model, both numerically (Section 3) and 
analytically (Section 4). As long as the function 
$f(x)$ is monotonically increasing and unbounded 
the qualitative behavior is the same: \\
{$\bullet \;\; $} Inflation is nearly de Sitter for 
a calculable period; and then \\
{$\bullet \; $} The Ricci scalar oscillates about 
zero with a calculable constant period and an 
amplitude that falls off like $t^{-1}$. \\
The universality of this behaviour was checked 
numerically by evolving functions $f(x)$ all the 
way from linear to exponential, and Section 4 
presents a derivation from the effective field 
equations.

Of course (single) scalar-driven inflation contains 
a free function, the scalar potential $V(\varphi)$, 
which can be fine-tuned to support a wide variety of 
expansion histories $a(t)$. However, the generic 
evolution of our model involves two distinct features 
which single-scalar inflation can never reproduce: \\
{$\bullet \; $} During the oscillatory phase, 
the Hubble parameter $H(t)$ actually drops below 
zero for brief periods; and \\
{$\bullet \; $} The derivative of the Hubble 
parameter $\dot{H}(t)$ is positive for about 
half of the time during the phase of oscillations. \\
The first feature is conducive to rapid reheating, 
while the violation of the weak energy condition 
implicit in the second is the hallmark of a quantum 
effect \cite{OW}.

Prominent among the list of topics for future work 
is perturbations. We need to show that the dynamical 
scalar mode of our model releases the energy of 
oscillations into matter to reheat the universe. 
If this happens, then the quantum gravity sector 
will go quiescent during a long epoch of conventional 
radiation domination. The subsequent transition to 
matter domination might even give rise to something 
like the current phase of acceleration, but this
requires modifications of our {\it simple} ansatz
which shall be described elsewhere \cite{NctRpw8}. 

Another important topic for future work is to 
derive and solve the equation for scalar 
perturbations, at least enough to compute the 
scalar power spectrum. One also needs that 
there be no long-range scalar force at late
times. Moreover, the equation for tensor 
perturbations is unchanged by our ``perfect 
fluid'' model. We need only use the expansion 
history $a(t)$ predicted by our model in order
to compute the tensor power spectrum.

\vspace{1cm}

\centerline{\bf Acknowledgements}
This work was partially supported by the European 
Union grant FP-7-REGPOT-2008-1-CreteHEPCosmo-228644, 
by the NSF grant PHY-0653085, and by the Institute 
for Fundamental Theory at the University of Florida.

\newpage

\begin{figure}
\centerline{\epsfig{file=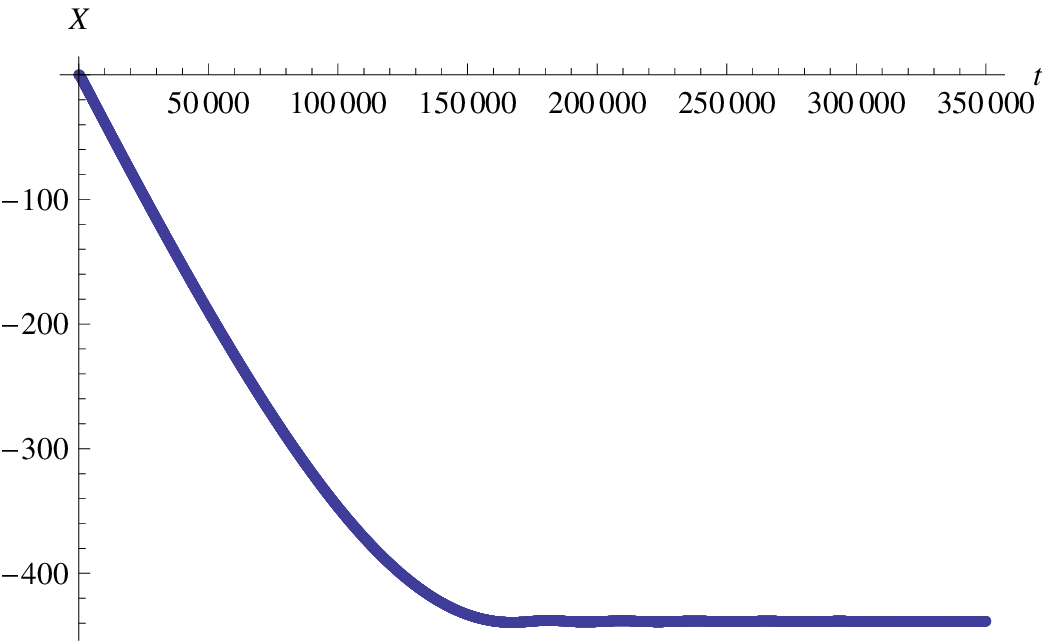,height=2.9in}}
\caption{\footnotesize The evolution of the source $X(t)$
over the full range for the exponential model.}
\end{figure}

\begin{figure}
\centerline{\epsfig{file=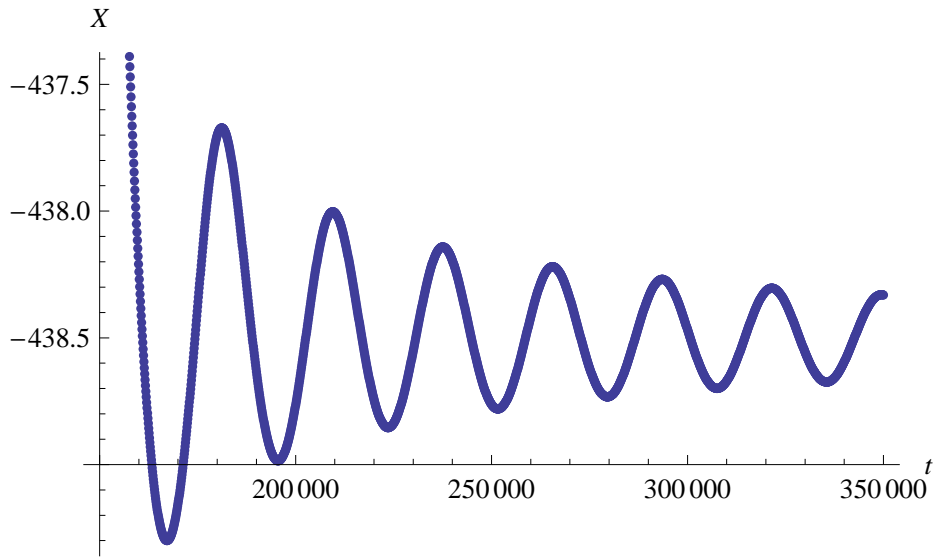,height=2.9in}}
\caption{\footnotesize The evolution of the source $X(t)$
during the oscillatory regime for the
\break \mbox{} \hspace{2cm}
exponential model.}
\end{figure}

\newpage

\begin{figure}
\centerline{\epsfig{file=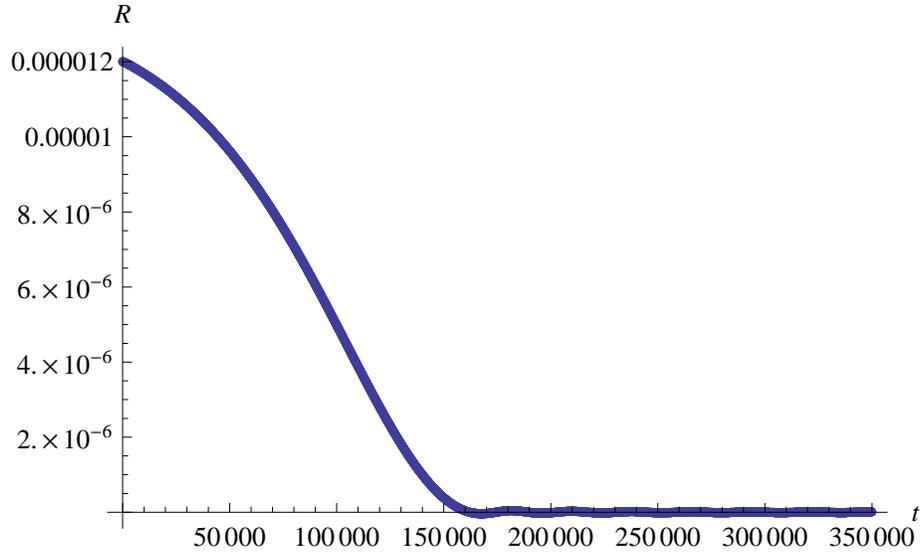,height=2.9in}}
\caption{\footnotesize The evolution of the curvature
scalar $R(t)$ over the full range for the
\break \mbox{} \hspace{2cm}
exponential model.}
\end{figure}

\begin{figure}
\centerline{\epsfig{file=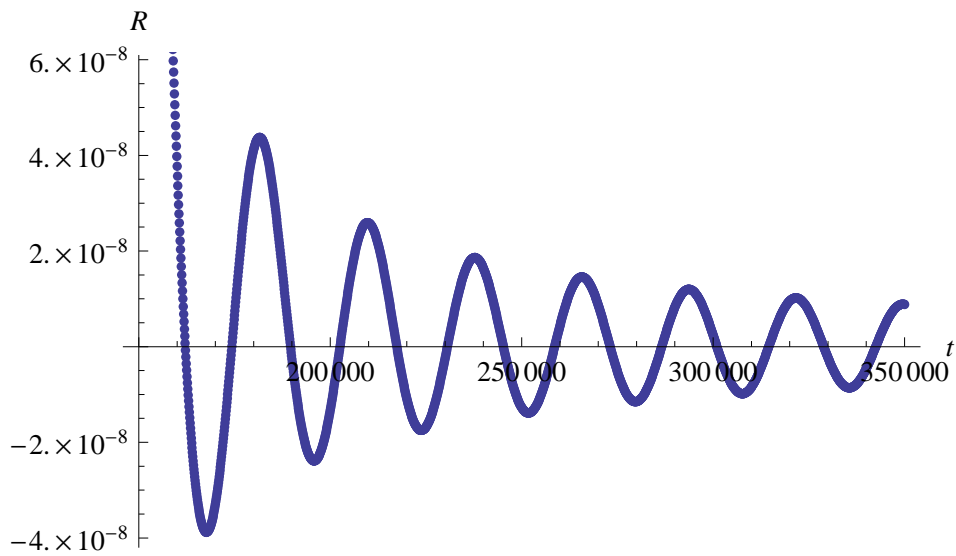,height=2.9in}}
\caption{\footnotesize The evolution of the curvature
scalar $R(t)$ during the oscillatory regime
\break \mbox{} \hspace{1.9cm}
for the exponential model.}
\end{figure}

\newpage

\begin{figure}
\centerline{\epsfig{file=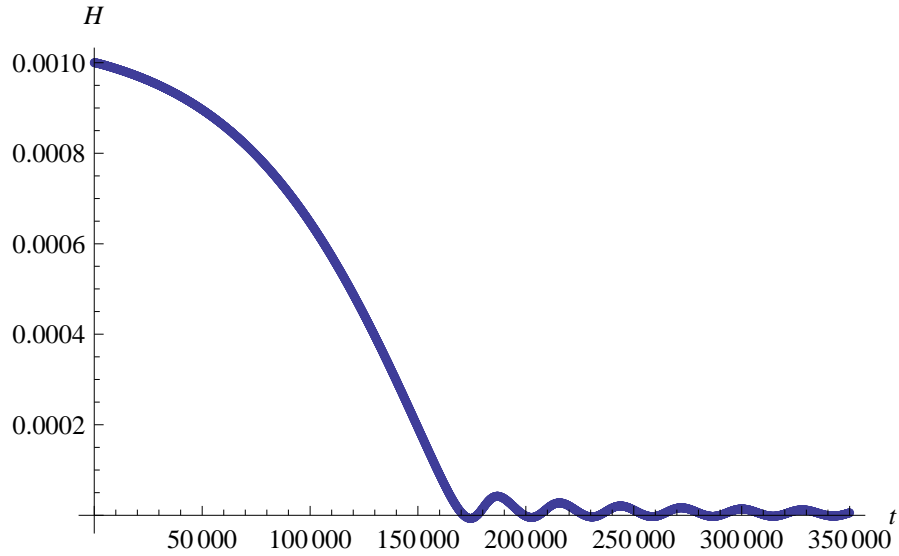,height=2.9in}}
\caption{\footnotesize The evolution of the Hubble
parameter $H(t)$ over the full range for the
\break \mbox{} \hspace{1.9cm}
exponential model.}
\end{figure}

\begin{figure}
\centerline{\epsfig{file=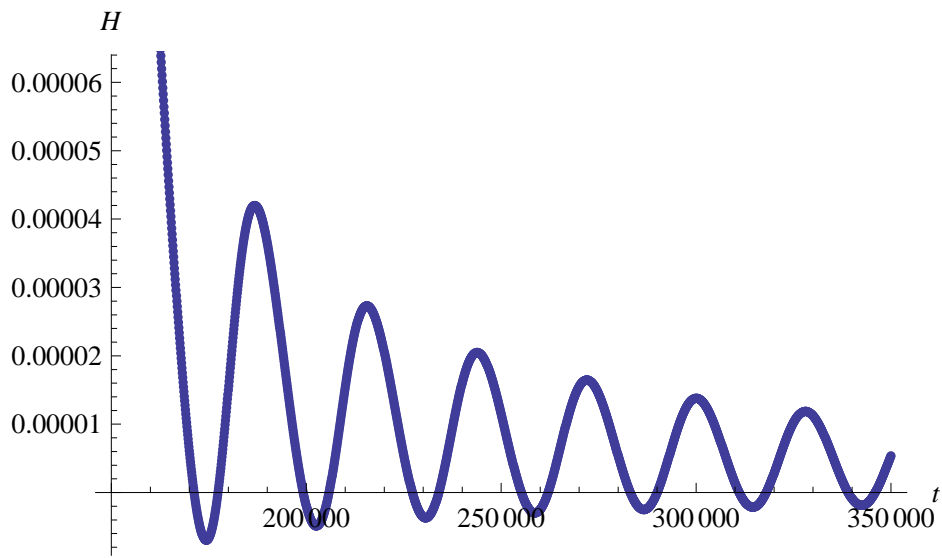,height=2.9in}}
\caption{\footnotesize The evolution of the Hubble
parameter $H(t)$ during the oscillatory regime
\break \mbox{} \hspace{1.8cm}
for the exponential model.}
\end{figure}

\newpage

\begin{figure}
\centerline{\epsfig{file=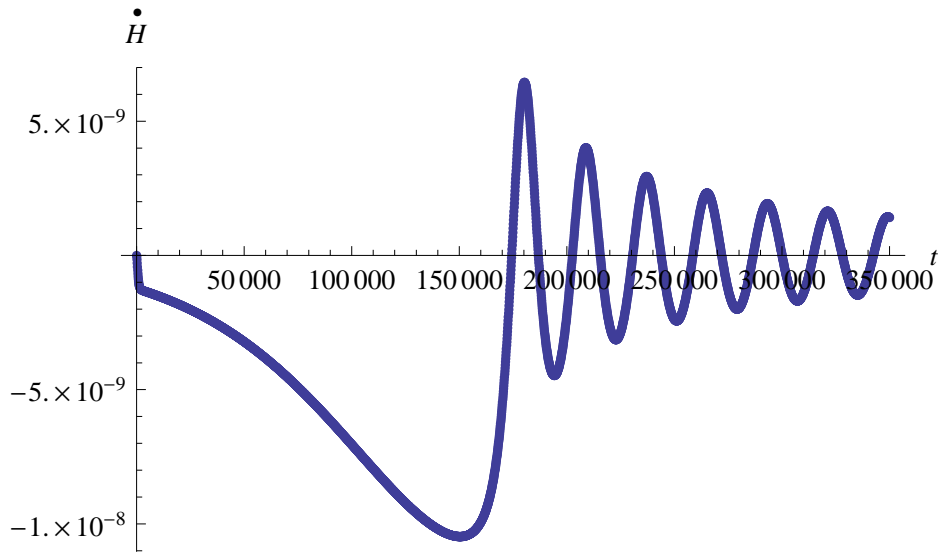,height=2.9in}}
\caption{\footnotesize The evolution of $\dot{H}(t)$ 
over the full range for the exponential model.}
\end{figure}

\begin{figure}
\centerline{\epsfig{file=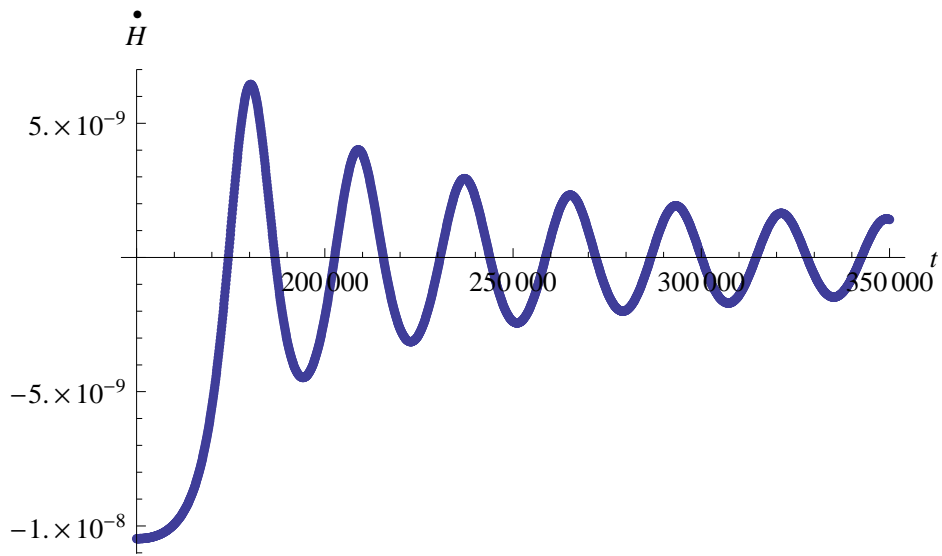,height=2.9in}}
\caption{\footnotesize The evolution of $\dot{H}(t)$ 
during the oscillatory regime for the exponential
model.}
\end{figure}

\newpage

\begin{figure}
\centerline{\epsfig{file=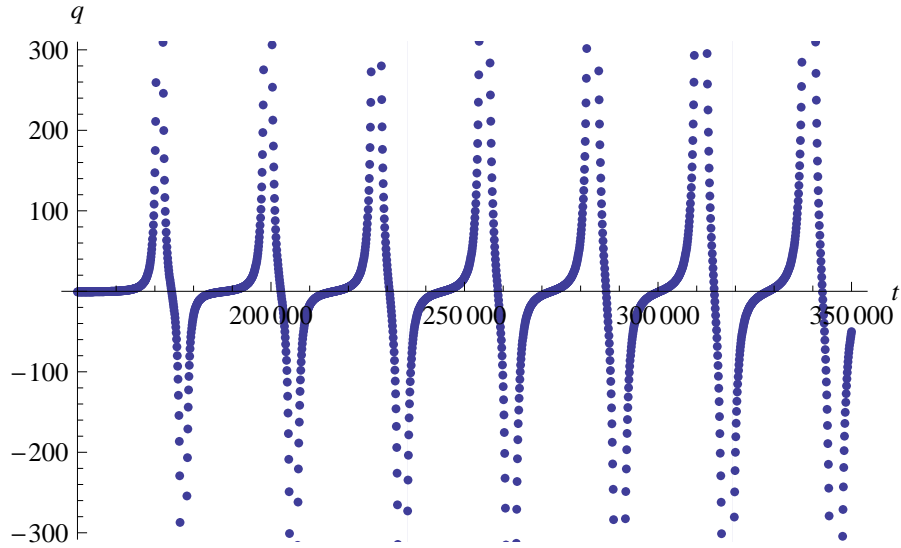,height=2.9in}}
\caption{\footnotesize The evolution of the deceleration
parameter $q(t)$ during the oscillatory regime
\break \mbox{} \hspace{1.75cm}
for the exponential model.}
\end{figure}

\begin{figure}
\centerline{\epsfig{file=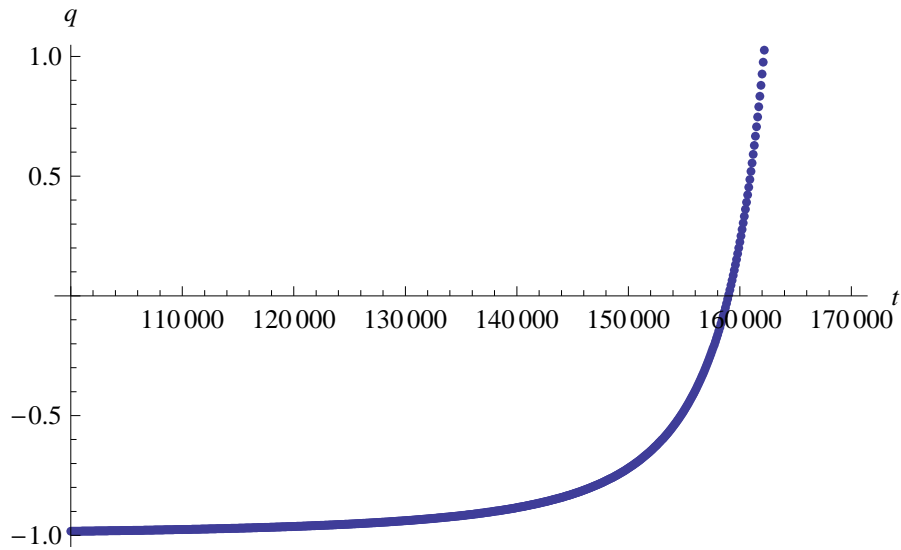,height=2.9in}}
\caption{\footnotesize The evolution of the decelaration
parameter $q(t)$ around the end of inflation
\break \mbox{} \hspace{1.95cm}
for the exponential model.}
\end{figure}

\newpage


\begin{figure}
\centerline{\epsfig{file=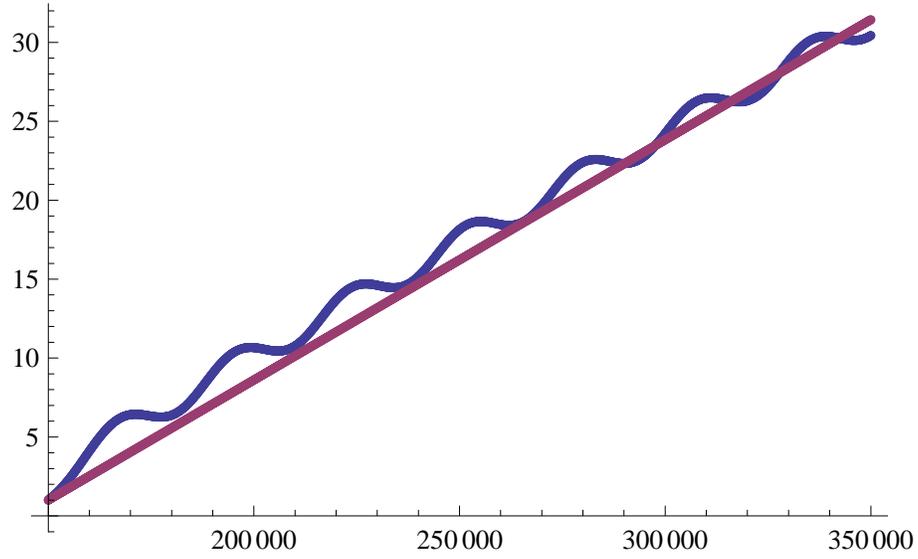,height=2.9in}}
\caption{\footnotesize The evolution of the scale 
factor ratio $[a(t)/a(150000)]$ during the oscillatory
\break \mbox{} \hspace{1.95cm}
regime for the exponential model {\it versus} a linear 
interpolation.}
\end{figure}


\begin{figure}
\centerline{\epsfig{file=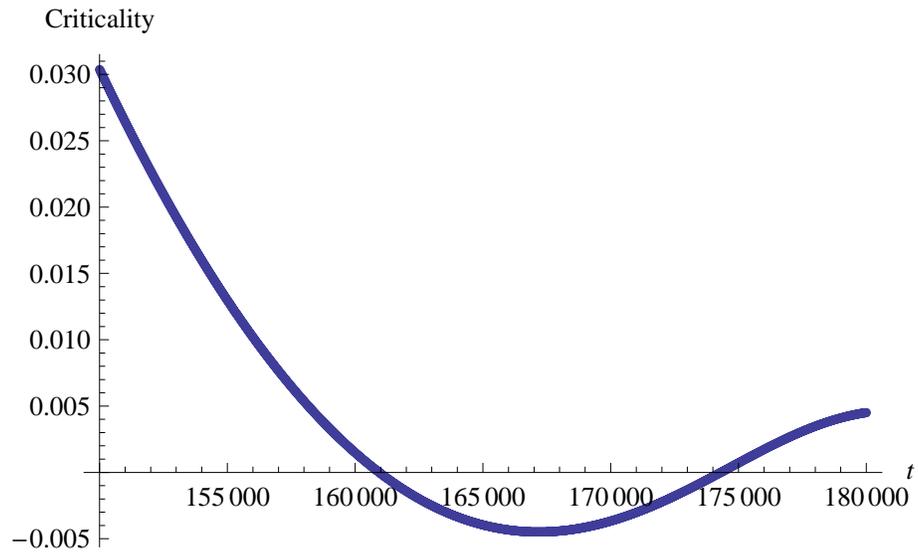,height=2.9in}}
\caption{\footnotesize Determining the critical point
$1 - 8 \pi \epsilon f[- \epsilon X_i] = 0$ for the
exponential model.}
\end{figure}

\end{document}